\documentclass{aastex}  

\begin{document}  

\title{The Distance and Morphology of V723~Cassiopeiae \\
(Nova Cassiopeia 1995)}  

\author{J.~E.~Lyke \& R.~D.~Campbell}     
\affil{W.~M.~Keck Observatory, 65--1120 Mamalahoa Hwy., 
Kamuela, HI 96743}
\email{jlyke@keck.hawaii.edu, randyc@keck.hawaii.edu}  

\begin{abstract} 
We present spatially resolved infrared spectra of V723~Cas 
(Nova Cassiopeia 1995) obtained over four years with the integral
field spectrograph (IFS) OSIRIS on Keck II.  Also presented are one 
epoch of spatially unresolved spectra from the long slit spectrograph 
NIRSPEC.  
The OSIRIS observations made use of the laser guide star adaptive
optics facility that produced diffraction limited
spatial resolution of the strong coronal emission features in the nova ejecta.
We remove the point--like continuum from V723~Cas data cubes
to reveal details of the extended nebula and find that emission
due to [\ion{Si}{6}] and [\ion{Ca}{8}] has an equatorial ring
structure with polar nodules; a strikingly different morphology than 
emission due to [\ion{Al}{9}], which appears as a prolate spheroid.  
The contrast in structure may indicate separate ejection events.  
Using the angular expansion and Doppler velocities observed over four 
epochs spaced
at one year intervals, we determine the distance to V723~Cas to be
$3.85^{+0.23}_{-0.21}~$kpc.  We present the OSIRIS three dimensional 
data here in many ways:  as narrowband images, one-- and 
two--dimensional spectra, and a volume 
rendering that reveals the true shape of the ejecta.
\end{abstract}  


\section{Introduction}  

Classical novae (CN) are eruptive variables resulting from
a thermonuclear runaway (TNR) on the surface of a white dwarf (WD). The
hydrogen--rich material that fuels the TNR is accreted onto the WD via
a disk of material, the source of which is a Roche--lobe filling
secondary star.  The TNR causes  $\sim10^{-4}~M_{\odot}$ of material
to be ejected at speeds of a few $\times10^{2}$ to a few
$\times10^{3}~km~s^{-1}$.  The ejection of material results in the CN
brightening by factors $\sim10^{4}$.  CN go through several stages in
the time following eruption \citep[see][for a description of CN
stages]{gehrz98}, one of which is the nebular
stage when CN become emission line sources.  The ejecta expand into a 
nebular remnant and as the continuum contribution declines,
forbidden lines often become dominant. Emission lines from CN often 
exhibit complicated
velocity structure that implies non--uniform ejecta \citep{hutchings72}.  
This non--uniformity is invoked by models to explain the
co--existence of lines arising from the recombination of hydrogen and
highly ionized forbidden transitions of other species. 

The study of CN nebular remnants can contribute to the understanding
of the CN process, the physics of the WD, shaping mechanisms from the TNR, 
winds, and binary motion, and the CN contribution to the chemistry
of the ISM.  Most CN become faint in the
nebular stage so the study of resolved novae ejecta is
relatively difficult.  HST studies have resolved ejecta in both the
visible and infrared (IR) 
\citep[for examples see][]{krautter02, paresce95, harman03}. 
However, since the CN often emit in exotic forbidden 
lines, direct imaging in the usual narrowband filters may not be the 
ideal method of study in some cases.  The high spatial resolution with 
adaptive optics from large ground
based telescopes has the capability to spatially resolve CN ejecta
early in the 
nebular phase and the nebular regions can be discerned precisely in
the bands of emission with an integral field spectrograph
(IFS).  This paper describes the use of OSIRIS (OH--Suppressing
Infrared Imaging Spectrograph; \citealt{larkin06}), an IR IFS
mounted on the Keck~II telescope equipped with laser guide star
adaptive optics (LGSAO); \citealt{wiz06}, to study the expanding nova shell of 
\object[Nova Cas 1995]{V723~Cas (Nova Cas 1995)}. 

V723~Cas was discovered by Yamamoto
on 24.5~August~1995, JD~2,449,954 \citep{hirosawa95}, and
spectroscopically confirmed by Iijima and Rosino \citep{ohshima95}.
Since then, its development has been followed closely as one of the slower, 
if not the slowest photometric developments known of classical novae 
\citep{iijima06, chochol97}. Remaining relatively bright, it has been widely
studied in many wavelengths and with many facilities \citep{rudy02,
evans03, heywood05}.  It became a super soft x--ray source \citep{ness08}
and has continued to exhibit strong emission lines in the
optical/IR. The angular extent of V723~Cas was not
sufficient for \citet{krautter02} using HST+NICMOS to detect extended 
emission at the time of their observations. The observations presented
here were made more than 7 years after those by \citet{krautter02} and
confirm detection of emission from the nova shell in various spectral
features.  The data presented here demonstrate the
capability of OSIRIS coupled to the Keck LGSAO system to resolve 
spatial details of a nova shell when its extent is on the order 
of $0\farcs25$ and to spectroscopically resolve velocity structure of
a nova shell expanding at approximately $250~km~s^{-1}$. 

In \S\ref{obs} we document the multi--epoch, multi--instrument 
observations of the expanding nova shell, in \S\ref{redux} we
explain the reduction 
process used on the IFS/AO data, in \S\ref{results}, we describe
the morphology of
the nova shell that is different among various emission lines and the 
expansion parallax method that employs knowledge of the shell shape 
to determine the distance to V723~Cas, and in \S\ref{discussion}, 
we suggest scenarios that may 
describe the origin of the shell morphology.  

\section{Observations}  
\label{obs}

V723~Cas, coordinates: 01$^{h}$05$^{m}$05$\fs$37  
+54$\arcdeg$00$\arcmin$40$\farcs$5 (J2000.0) was observed as part of 
survey of recent novae on 2004 August 22 UT using NIRSPEC (Near
Infrared Spectrometer) on Keck~II
in the low resolution mode with the  42$\arcsec\times0\farcs$38
long--slit \cite{mclean98}.  The telescope was dithered in an ABBA
pattern, the preferred method to aid in background subtraction 
\cite{gehrz92}.  An A0V star 
HD~6313 was used to correct for telluric absorption and to serve as 
a rough flux calibrator.  The seeing on these nights was about
0$\farcs$7; therefore, slit losses render flux calibrations 
good to only 20--30$\%$.  

The NIRSPEC spectra of V723~Cas showed strong H and coronal emission 
features nearly 10 years after outburst (see Figure~\ref{fig:1d}).  Based upon 
the observed expansion velocities of the emission lines, the time
since outburst, and the estimated distance to V723~Cas, we believed 
V723~Cas would be an ideal object to obtain spatially resolved spectra with 
OSIRIS and the Keck~II laser guide star adaptive optics (LGSAO).  

V723~Cas was observed with OSIRIS on 2005 September 13, 2006 August
31, 2007 September 03 UT, and 2008 August 09 UT.  A summary of our 
observations is presented 
in Table~\ref{tab:details}.  OSIRIS uses a lenslet array in the focal 
plane to focus incident light into a pupil plane. Each pupil is then 
dispersed by a diffraction grating and focused onto a detector 
\citep{larkin06}.  A 
dedicated data reduction pipeline (DRP) extracts the two--dimensional 
(2D) spectra from the detector and produces a three--dimensional (3D) 
data cube with two spatial dimensions ($x$, $y$) and one wavelength 
dimension ($\lambda$; \citealt{krabbe04}).  We describe this process 
in more detail in \S\ref{redux}.  The Keck LGSAO system requires a 
natural tip--tilt guide star and in this case, V723~Cas was its own tip--tilt 
guide star.  In 2005, V723~Cas was $R~=~14.8~mag$ and by
2008 had faded to $R~=~15.8~mag$.  
On all nights, we used the $0\farcs035$ per lenslet plate scale and 
moderate--band filters to observe an approximate field of view of 
$1\farcs5\times2\farcs24$.  Broadband filters would cover the entire 
$K$--band, but reduce the field of view by a factor of 3.  
Our integrations were 900~seconds and the 
telescope was dithered to measure the sky background.  When additional
signal was required, we observed additional nod pairs. For 2005 
September 13, we observed in the Kn1, Kn2, and Kn3 moderate--bandwidth 
filters yielding wavelength coverages of $\lambda =
1.955-2.055~\micron$, $\lambda = 2.036-2.141~\micron$, and 
$\lambda = 2.121-2.229~\micron$ respectively.  
Conditions on 2005 September 13 were 
marginal:  $K$--band seeing was $0\farcs7$ and relative humidity 
was above 50$\%$.  The LGSAO system performed well giving 
a $K$--band full--width, half maximum (FWHM) of $0\farcs050$.  We estimated
the Strehl ratio to be 20$\%$. For 2006 August 31, we observed in the 
Kn1 and Kn3 bands yielding wavelength coverage of $\lambda = 
1.955-2.055~\micron$ and $\lambda = 2.121-2.229~\micron$.  The 
measured $R$--band seeing on 2006 August 31 was $0\farcs65$.  There 
were patchy clouds for part of the night, but none were observed in 
the location of V723~Cas during our observations.  The Keck~II LGSAO 
system employs redundant systems for safety and performance 
monitoring that can double as cloud detectors. The first is laser 
safety observers who are posted outside the observatory to watch for 
airplanes.  Their job is to shutter the laser if an airplane strays 
too close to the beam. Additionally, they are instructed to shutter 
the laser if they see laser scatter from thin clouds to prevent 
additional sky brightness that would affect the other Mauna Kea 
observatories.  The second is a photometrically calibrated tip--tilt 
sensor.  The laser safety observers reported no laser scatter and 
the measured $R$--band magnitude of V723~Cas did not vary during 
our observations. The LGSAO system produced a near--diffraction 
limited image in $K$--band FWHM of $0\farcs050$.  Sky conditions for 
our 2007 September 03 observation were excellent: clear skies, low 
humidity, and $K$--band seeing of $0\farcs3$.  Furthermore, this 
observation was obtained after a wavefront controller and sensor 
upgrade to the Keck AO system.    
We observed in the Kn1, Kn3, and Kn5 bands yielding wavelength 
coverage of $\lambda = 1.955-2.055~\micron$, $\lambda = 
2.121-2.229~\micron$, and $\lambda = 2.292-2.408~\micron$ 
respectively.  For our 2008 observations, the sky was clear and 
$R$--band seeing was measured to be $0\farcs6$.  We observed in the 
Kn1, Kn3, and Kn5 bands as in 2007.  During each 
night, we observed an A0V star in the same instrument configurations 
for telluric correction.      

\section{Data Reduction} 
\label{redux}  

\subsection{NIRSPEC} 
The NIRSPEC spectra are conventional long--slit 
spectrograph data. Calibration frames were acquired at various times 
during the night using the internal flat 
lamp and arc lamps and with the same instrumental setup as our 
science observations.  We used these calibration frames and the
IDL--based reduction package 
REDSPEC\footnote{http://www2.keck.hawaii.edu/inst/nirspec/redspec.html}
to reduce the NIRSPEC data in the standard way.     

\subsection{OSIRIS} 
The OSIRIS DRP is an IDL--based package built on separate modules and is 
available for download from the Keck Observatory OSIRIS Tools 
Page\footnote{http://www2.keck.hawaii.edu/inst/osiris/tools/}. 
The DRP calls the modules in sequential order to reduce raw, 2D images
into 3D data cubes.  After this basic reduction, there are DRP modules
that mosaic data cubes and assist in telluric correction.  For 
telluric correction, modules extract one dimensional (1D) spectra 
of bright objects, remove intrinsic spectral features from 1D 
spectra, remove a blackbody curve from 1D spectra, and divide 3D 
data cubes by 1D spectra.  When the 2D pixel information is 
converted into a 3D data cube, each spatial $x,y$ position in the 
data cube is linked to a spectrum and not just an intensity.  We follow the 
Euro3D convention and refer to each $x,y$ position in the 
3D data cube as a "spaxel", or spatial pixel.  Each spaxel maps to a 
lenslet so here, each spaxel is $0\farcs035$ per side.

\subsubsection{Basic DRP} 
The basic reduction sequence for these data removes detector artifacts as 
follows: 1) pairwise sky subtraction; 2) removal of crosstalk associated 
with bright spectra on a single row of the detector; 3) adjustment of the 
32 detector channels to remove systematic bias; 4) rejection of electronic 
glitches from detector readout; and 5) rejection of cosmic rays. After 
these steps, the OSIRIS DRP reconstructs a data cube by extracting 2D 
spectra from the raw frame.  Extraction requires mapping the PSF of each 
lenslet position as a function of wavelength with a white--light source.  
The OSIRIS DRP iteratively assigns flux from each pixel to its 
corresponding lenslet spectrum using the PSF maps.  Once this is done, 
each spectrum is resampled onto a linear wavelength grid via interpolation 
and a wavelength solution to spectral arc lines.  The final step inserts 
the extracted spectra into their proper spatial locations in a data cube.  
OSIRIS data cubes are linear in spatial $x$ and $y$ and wavelength channel 
$\lambda$.  In $K$--band, each wavelength channel is separated by 
$0.00025~\micron$.

\subsubsection{Telluric Correction} 
After the object and telluric star data are reduced to the basic cube
level, additional steps must be taken before the data can be
analyzed. First, one extracts a 1D spectrum of the telluric star from
its cube; with intrinsic features removed.  For slit spectrograph
data, the next step is to divide the object spectrum by the
telluric spectrum.  The data cube analog is to divide the
spectrum of each lenslet by the 1D telluric spectrum.  If not done carefully, 
this approach
can lead to poor telluric correction.  A diagnostic for the quality of 
telluric correction comes from blank sky regions in the data
cube: background--subtracted blank sky
should have a flat spectrum both before and after telluric correction.
When one has poor telluric correction, one sees over-- or
under--corrections of the atmospheric absorption features in blank sky
regions.  We have found that blank sky regions of pairwise--subtracted
data cubes often have a constant, non--zero offset within each channel
(wavelength slice) that is well--modeled by a constant across wavelengths.  We
attribute this offset to the background level changing between the
900~second object frame and the 900~second sky frame.  After
subtracting the measured constant from each channel, we divide the
spectrum of each lenslet in the object cube by the 1D telluric
spectrum and see the expected diagnostic results:  a slightly noisier 
continuum with values similar to the continua on either side of the 
telluric feature.  

\subsubsection{Continuum Subtraction}  

V723~Cas proved to be a bright continuum source throughout our
observations.  We do not detect continuum emission in the
nebular region.  The continuum of a classical nova long after outburst
is dominated by light from the accretion disk as evidenced by studies 
of eclipsing systems (Horne 1985).  As such, it is
not resolved and thus represents the PSF of the data. In order to
better study the extended emission line regions, we removed the
point--source continuum from the data cube.  The ability to separate the
3D data cube into a 2D image at each wavelength makes the continuum 
removal process straightforward.  1) The integrated spectrum of the 
source is examined to determine a spectral range
over which there are no emission lines and no atmospheric features.  
2) A continuum image is created 
by median--combining the pure continuum wavelength channels.  Note
that this continuum image has the same $x$ and $y$ dimensions as the
data cube.  3) A 1D spectrum is extracted from a $3\times3$ spaxel
region of the object data cube centered on the peak 
of the continuum image.  The 9--spaxel extraction gives both a
high signal--to--noise ratio and averages out the impact of a bad
data cube element.  4)
An interpolation is performed over the emission features in the 
extracted spectrum
and 5) the emission line--free continuum is smoothed with a boxcar 5 
wavelength channels wide.  6) At each wavelength, the value of 
the modified continuum spectrum is used to scale the peak of the 
source in the continuum image.  The scaled continuum image is
subtracted from each wavelength channel of 
the data cube.  The result is demonstrated in Figure~\ref{fig:2d} as the 
2D spectrum with the continuum removed accentuates the extended emission 
and shows little residual emission at the position of the point--like 
continuum.  The subtraction process does not require PSF modeling or 
fitting and 
is therefore robust. We note however that as V723~Cas fades, the continuum 
subtraction residuals are larger relative to the shell emission in our 
later epoch observations.    

\section{Results}  
\label{results}

\subsection{NIRSPEC} 
The NIRSPEC spectrum of V723~Cas shows that it is a strong emission
line source, some 3285 days after discovery.  As Figure~\ref{fig:1d}
shows, the most prominent lines are Pa$\alpha$,
[\ion{Si}{6}]~1.96~$\micron$, [\ion{Al}{9}]~2.04~$\micron$, Br$\gamma$,
[\ion{Ca}{8}]~2.32~$\micron$, and [\ion{Si}{7}]~2.48~$\micron$.  
The list of identified lines appears in
Table~\ref{tab:linelist}.  Nearly all lines are double--peaked, indicating 
expansion as expected.  The brightest emission lines were targeted for 
OSIRIS follow--up.  

\subsection{OSIRIS} 
\begin{figure}
\plotone{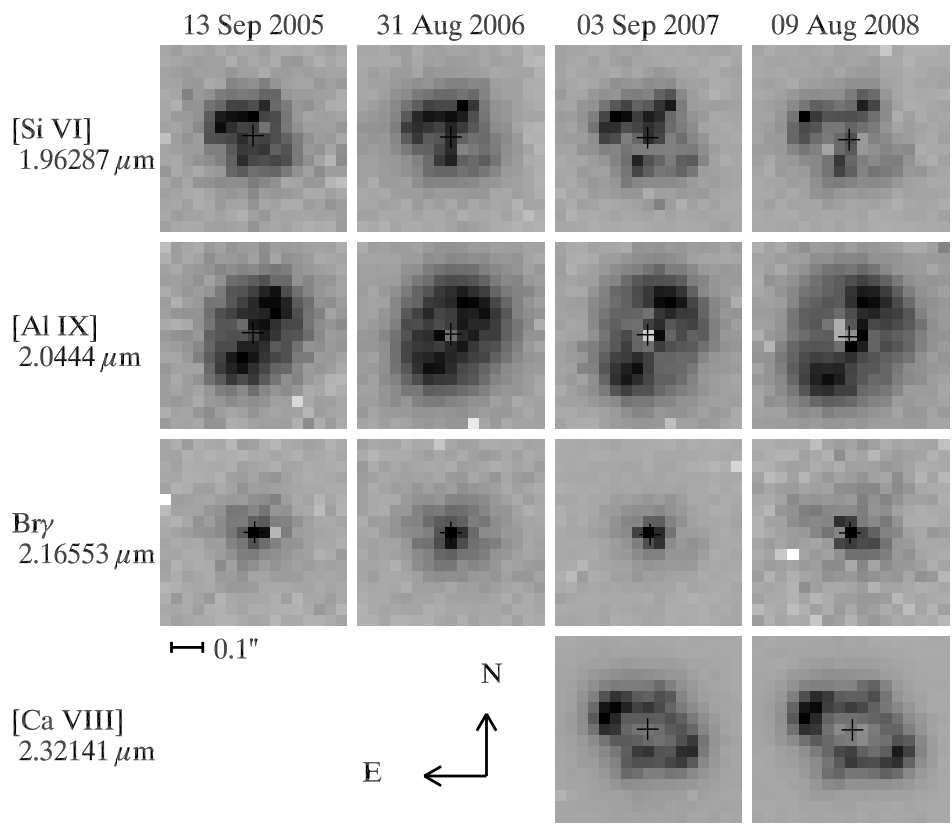}
\caption{Narrowband images of V723~Cas of various emission lines show
different morphologies.  Each row holds images from one species and
each column contains images for one epoch.  The continuum has been
subtracted from each image.  Images are normalized but each image has
its own linear stretch to accentuate features.  The passbands for the
[\ion{Si}{6}], [\ion{Al}{9}], Br$\gamma$, and [\ion{Ca}{8}] images are
$0.0055~\micron$, $0.00775~\micron$, $0.0055~\micron$, and 
$0.00725~\micron$ respectively.\label{fig:nb}}
\end{figure}

The nova shell of V723~Cas is resolved in both the spatial
and wavelength dimensions of the 3D data cube. The high spatial resolution 
data reveal a complex structure of the nova ejecta. In order to highlight 
certain features of the nova ejecta, we display the data in various
forms.  These include 1D spectroscopy (Figure~\ref{fig:1d}), 
2D spectroscopy (Figure~\ref{fig:2d}), narrowband imaging 
(Figure~\ref{fig:nb}), and 3D spatial--velocity
(Figure~\ref{fig:face} and Figure~\ref{fig:edge}).   

\subsubsection{1D Spectroscopy}
\label{sec:1d}
\begin{figure}
\plotone{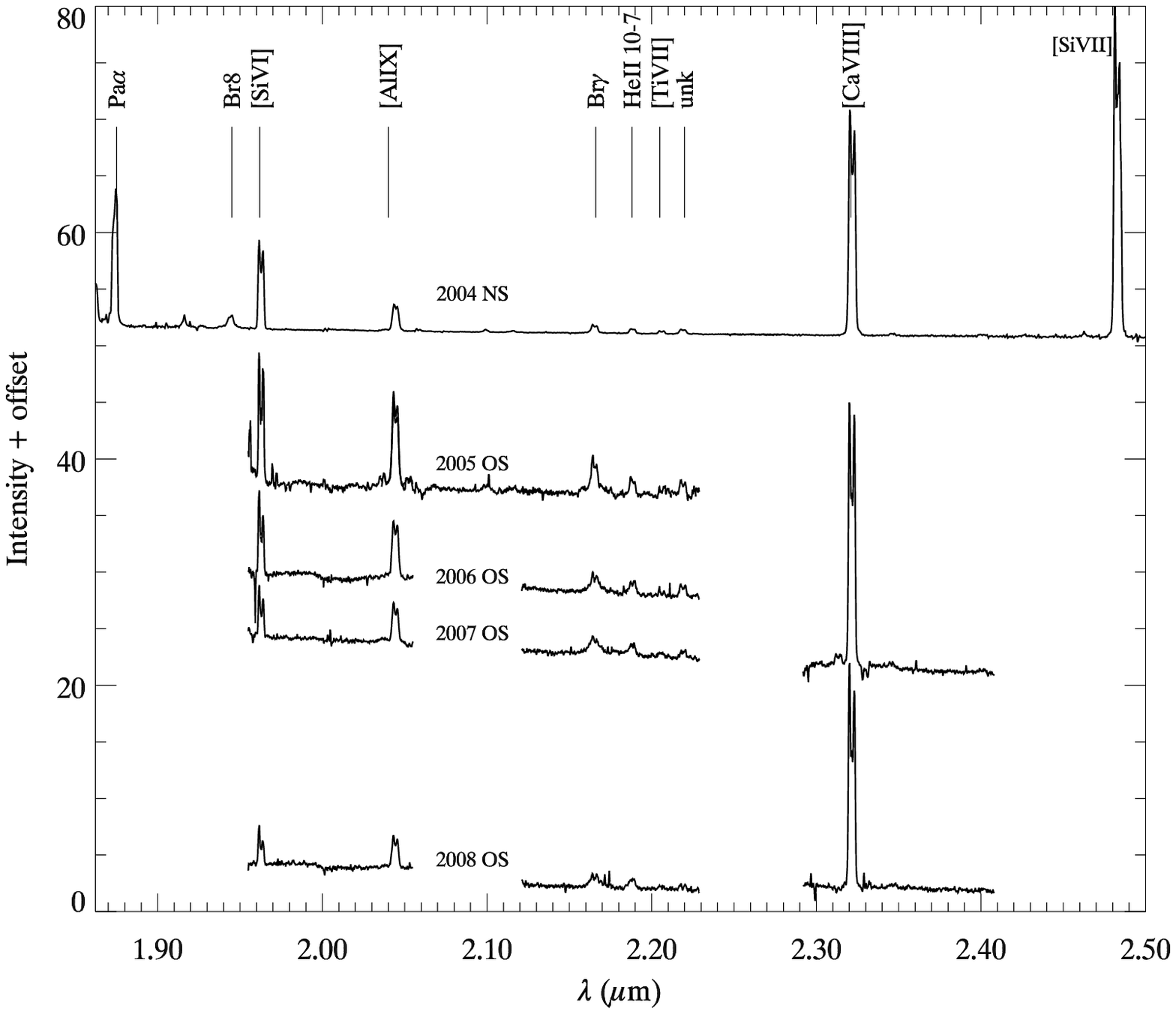}
\caption{One dimensional spectra for NIRSPEC (NS) and OSIRIS (OS) from
2004 through 2008.  We note that the [\ion{Al}{9}] feature gets stronger 
as compared to the [\ion{Si}{6}] feature with time.  Relative flux ratios
are given in Table~\ref{tab:linelist}.\label{fig:1d}}
\end{figure}

The OSIRIS data cube can be integrated over both spatial dimensions to 
create the equivalent of a slit spectrum of V723~Cas.  This allows a 
direct comparison
with the NIRSPEC spectrum as shown in Figure~\ref{fig:1d} and 
Table~\ref{tab:linelist}.  For Figure~\ref{fig:1d} we extracted an 
$11~\times~11$ spaxel box centered on the continuum source to approximate 
the $0\farcs38$ slit used in the NIRSPEC spectrum.  It is evident 
that [\ion{Al}{9}] is
stronger compared to [\ion{Si}{6}] in 2005 than in 2004.  Rudy et al. (2002) 
detected a weak [\ion{Al}{9}] line with a flux ratio compared to that of 
[\ion{Si}{6}] of $0.002~\pm~0.0015$ in 1999 and $0.005~\pm~0.0015$ 
 in 2000.  Thus, over the 1999--2008 span of observations the 
[\ion{Al}{9}] has strengthened
significantly each year with respect to [\ion{Si}{6}].  
The OSIRIS spectral coverage is limited compared to that 
of NIRSPEC so certain lines, such as [\ion{Si}{7}]~2.48~$\micron$ and 
Br$\delta$ are not included in our spatially resolved study.  The spectra 
from the OSIRIS Kn3 filter and NIRSPEC did include features seen in 
\citet{rudy02}, including \ion{He}{2} (10--7)~2.1882~$\micron$, 
[\ion{Ti}{7}]~2.2050~$\micron$, and an unidentified line at near 
2.2188~$\micron$.  While \citet{rudy02} ultimately found this line to be 
unidentified, they did offer [\ion{Fe}{3}]~2.2178~$\micron$,
hereafter [\ion{Fe}{3}], as a possible 
source.  Emission from [\ion{Fe}{3}] does appear in OSIRIS data 
taken near the Galactic center (Ghez \& Do, private communication) and our 
analysis 
shows the unidentified line center to be 2.2190~$\micron$, a
difference of 5 wavelength channels.  For this reason, we concur 
with \citet{rudy02} that this line is not due to [\ion{Fe}{3}].  
An important result of the spectral analysis is that the
radial velocity of the nova shell remains constant over our 5 years of
observation (see Figure~\ref{fig:const_vel}).  We use this fact in 
\S\ref{distance}, to determine the distance to V723~Cas via expansion
parallax.  
    
\subsubsection{Imaging} 
A useful feature of integral field instrumentation is the capability
to create images using a customized bandpass tailored to any feature of 
the spectrum selected after the data have been acquired. Figure~\ref{fig:nb} 
shows custom narrowband images of
selected emission features in the resolved nova ejecta from the
coronal lines of [\ion{Si}{6}], [\ion{Al}{9}], [\ion{Ca}{8}] and the
hydrogen recombination line, Br$\gamma$.   Most noticeable is the
difference between the [\ion{Al}{9}] emission when compared to the
[\ion{Si}{6}] and [\ion{Ca}{8}] features. The
[\ion{Al}{9}] feature is relatively smooth and consistent with a
prolate spheroid shell. The other features have
an equatorial torus and polar nodules similar to that seen
in other novae ejecta such as DQ~Her \citep{mustel70}, HR~Del 
\citep{harman03},V1974~Cyg \citep{paresce95}, and FH~Ser \citep{gill00}. 
For [\ion{Si}{6}] and [\ion{Al}{9}], the ejecta expansion is evident 
over the four years of OSIRIS observations      

\subsubsection{2D Spectroscopy} 
\begin{figure}
\plotone{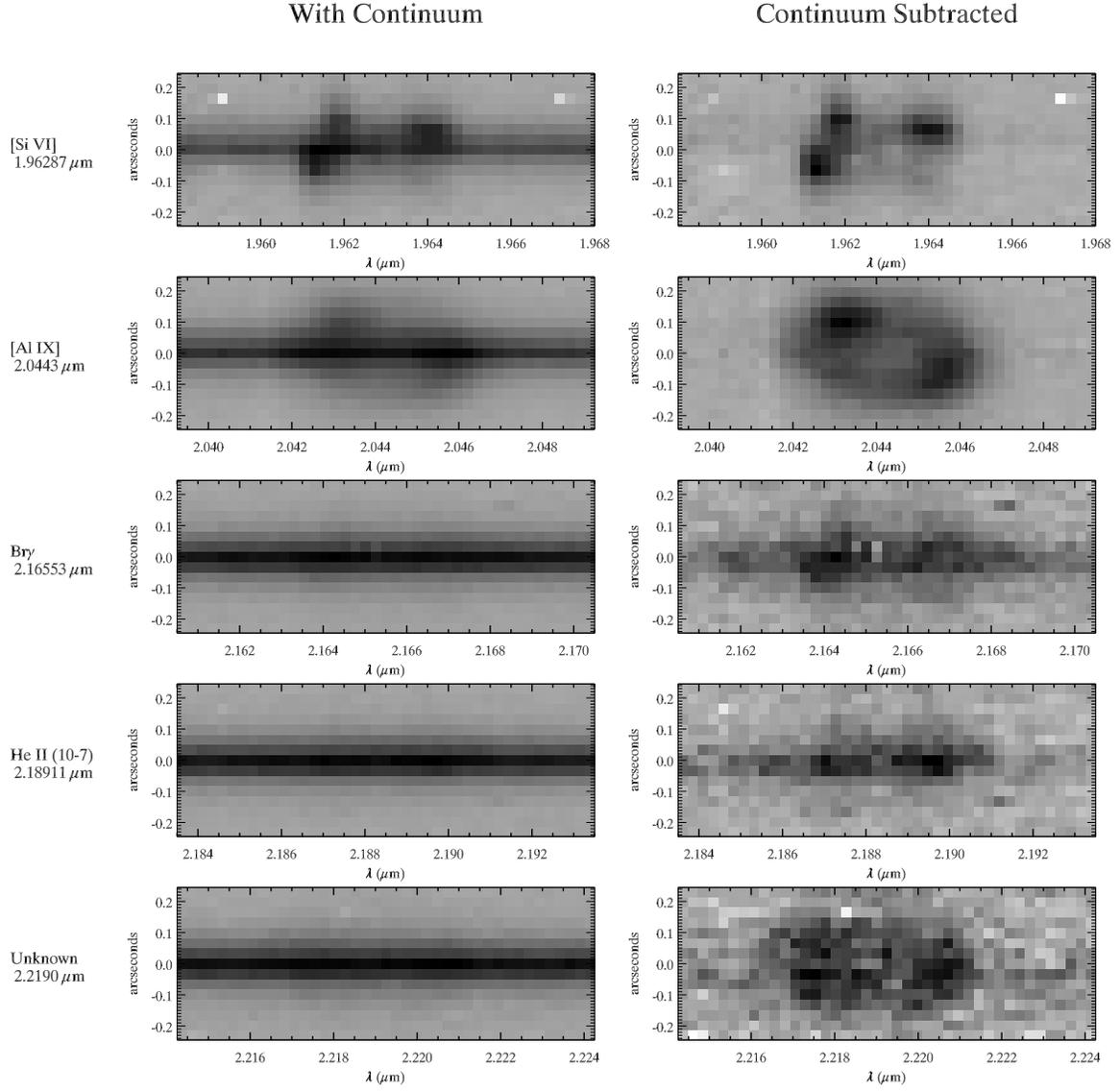}
\caption{Two dimensional OSIRIS spectra from 2006.  The left hand
column shows the original spectra, the right hand column shows the 
continuum--subtracted spectra  The bright [\ion{Si}{6}] and [\ion{Al}{9}]
lines show a striking difference in morphology.\label{fig:2d}}
\end{figure}

IFS data can be presented in a conventional 2D format for the purpose of 
highlighting certain features and to enable comparison to long slit data 
of more conventional spectrometers.  Figure~\ref{fig:2d} is a sampling of 
2D spectra of selected bands from the 2006 OSIRIS data.  The 2D spectra 
are produced by extracting a portion of the cube that would correspond to 
a slit--like region of interest if the slit were aligned north--south and 
$0\farcs5$ in width. The intensity is summed in the $x$ spatial dimension 
(E--W) and then displayed as a function of wavelength versus the $y$ spatial 
dimension (N--S).  In the left column of Figure~\ref{fig:2d}, the spectra 
are shown with the continuum.  In the right column, the point--like 
continuum has been subtracted.  It is apparent that the [\ion{Al}{9}] 
emission is distinctive in its spatial--velocity structure.  We note that 
the continuum subtracted data confirm that the coronal emission 
originates exclusively in the ejecta while the hydrogen recombination 
Br$\gamma$ line appears to come from both the ejecta and the central 
source.  More clearly evident in the 2D spectra than in narrowband 
images, the extended portion of Br$\gamma$ and \ion{He}{2}~(10--7) appear 
to have a structure similar to that of [\ion{Si}{6}] and [\ion{Ca}{8}].  
In contrast, the unknown feature centered at 2.2190~$\micron$ is more like 
[\ion{Al}{9}] than [\ion{Si}{6}].  

\subsubsection{Distance}
\label{distance} 
\begin{figure}
\plotone{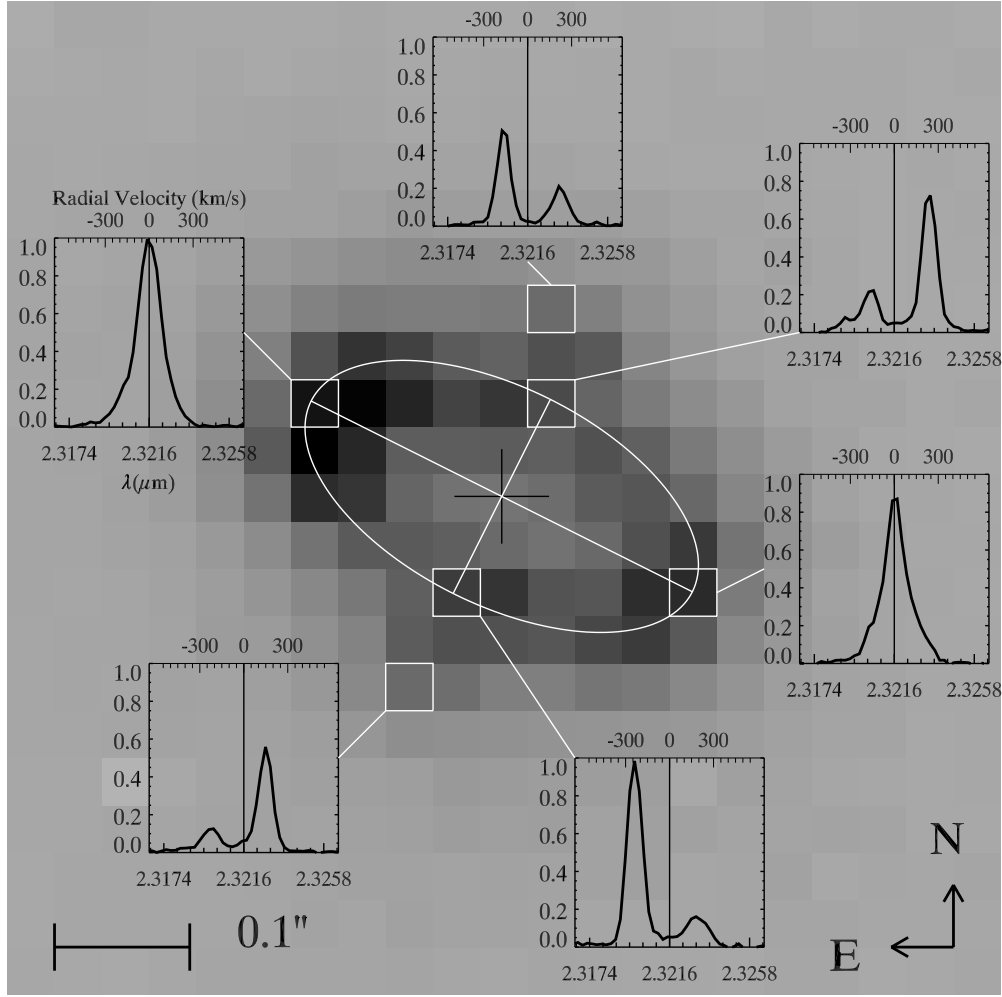}
\caption{Spectra extracted from individual spaxels are shown overlaid
on the [\ion{Ca}{8}] image from 2007.  This image is
identical to that shown in Figure~\ref{fig:nb}.  The ellipse shows 
a circular torus inclined to the
plane of the sky at 62$\degr$.  Spectra from the regions where
the ellipse intersects the major axis of the ellipse show no
net velocity.  Spectra from the regions where the ellipse 
intersects the minor axis of the ellipse show primarily blue--shifted
emission along the SE edge and primarily red--shifted emission along
the NW edge.  Emission from the regions above NW edge and below the SE
edge of the ellipse have the opposite velocity sense of the regions
along the minor axis of the ellipse.  This emission is attributed to
polar caps of emission.  The black "plus" sign denotes the location of
the subtracted continuum.\label{fig:inset}} 
\end{figure}

The spatially resolved IFS 3D data provide all the necessary information 
for an accurate distance determination. Angular expansion and radial 
velocity measurements can be derived from the same dataset, an advantage 
over methods that use different measurements acquired at different times. 
We demonstrate this in Figure~\ref{fig:inset}, where we have overplotted
spectra at various locations of the [\ion{Ca}{8}] feature from 2007.  The
3D data cube allows us to spectroscopically confirm that the apparent
ellipse of emission is consistent with an inclined circular torus with
polar nodules perpendicular to the torus.  We have overlaid an ellipse on 
the image to guide the reader's eye.  Spectra at positions along the major
axis of the torus have zero radial velocity; therefore, they are moving in
the plane of the sky.  At positions along the minor axis of the torus, the
spectra have double--peaked emission, with the SE portion more blueshifted
and the NW portion more redshifted.  The spectra at positions away from
the torus have the opposite velocity direction as the spectra on the torus,
indicating a polar nodule structure.  We make use of our knowledge of the 
morphology and inclination 
of the system to accurately determine a distance to V723~Cas. 
Rather than assume spherical symmetry of the nebula with Hubble flow 
properties, the distance analysis in this study uses the equatorial torus 
portion of the expanding ejecta and is based on the assumption that the 
torus is circular, or azimuthally symmetric.   

We create a zero velocity image for each emission feature by selecting the 
wavelength channel or channels that correspond to the central wavelength
of each emission feature.  \citet{reconditi93} give the wavelength of 
[\ion{Si}{6}] to be $1.96287~\pm~1.0~\micron$.
This corresponds to wavelength channels 31 and 32 in our Kn1 data
cubes.  We determine the angular expansion rate by measuring the centroids of
emission in the zero velocity image for [\ion{Si}{6}], or in other
words, the cross section of the torus that is expanding perpendicular
to our line of sight.  In a zero velocity image, the equatorial torus
of emission appears as two peaks on opposite sides of the continuum
location and moves purely in the plane of the sky (see
Figure~\ref{fig:expand}).  We use only [\ion{Si}{6}] because its
morphology lends itself to this analysis and the observations of
[\ion{Si}{6}] cover a four year span.  We do not include
[\ion{Ca}{8}] in the distance analysis because we have only two epochs
of observations and we do not assume that the two lines are perfectly 
coincident nor co--moving.  We fit the angular separation of the 
two emission peaks in the zero velocity image of 
each epoch by the method of linear least squares
to determine an angular expansion rate, 
$\phi~=~27.51~\pm~1.52~mas~yr^{-1}$ as seen in Figure~\ref{fig:expand}.  
We note that this determination does not depend on the assumed time of 
outburst (t$_{0}$).  From the zero velocity emission peaks, we measure
the position angle of the torus to be 63$\fdg7~\pm~$2\fdg0. 

\begin{figure}
\plotone{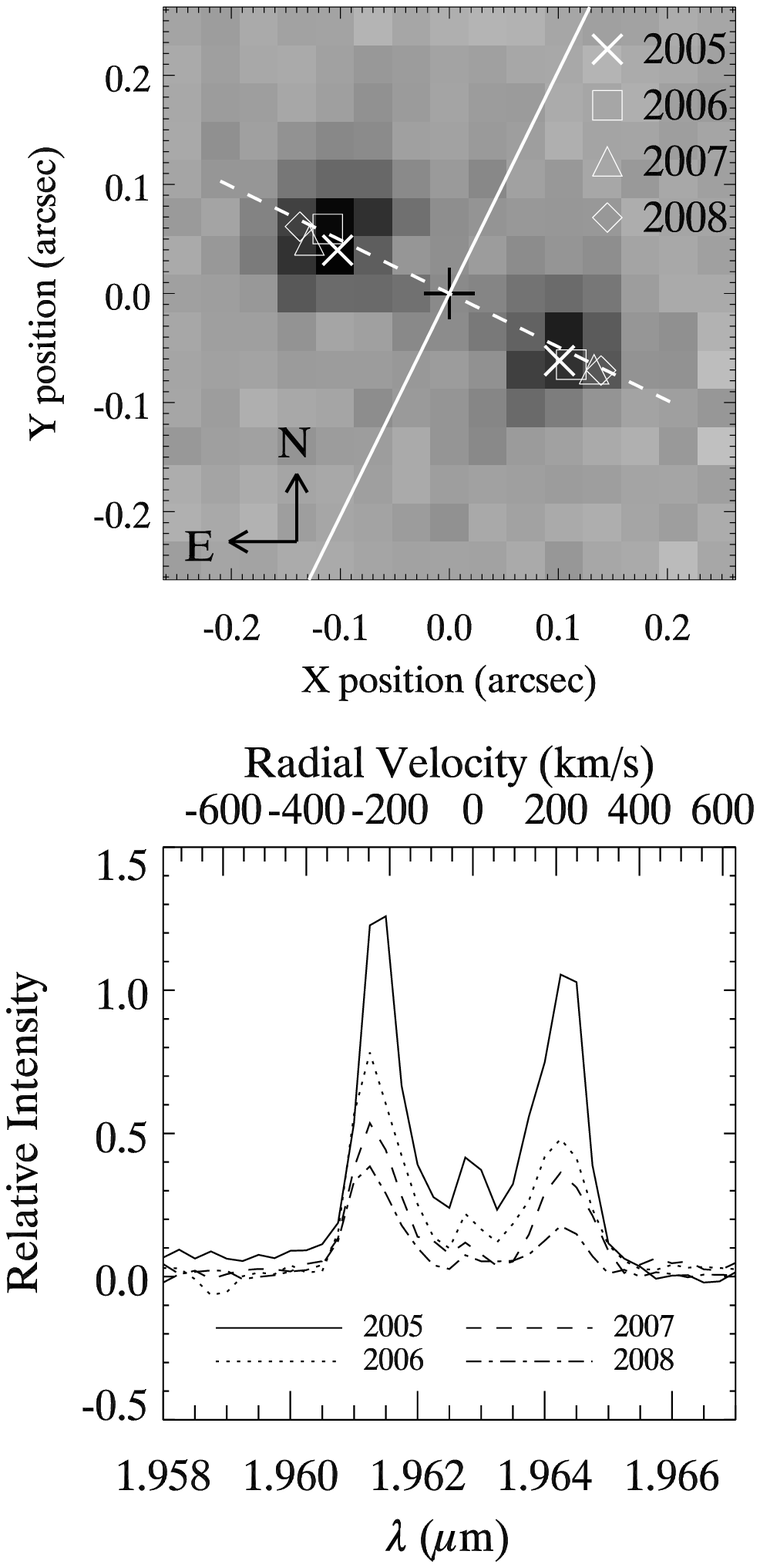}
\caption{Constant expansion velocity.  The upper panel shows the
centroid of emission at the zero velocity of [\ion{Si}{6}] for four epochs 
superimposed upon the [\ion{Si}{6}] image from 2005.  The passband of
the image is $0.00050~\micron$.  The black "plus" marks the position of
the subtracted continuum, the white dashed line has the slope of the
best fit line through the data points, and the white solid line is
perpendicular to the dashed line and shows the line along which the
spectra in the lower panel were measured.  The lower panel shows the
spectra of V723~Cas centered on the [\ion{Si}{6}] line.\label{fig:const_vel}} 
\end{figure}

\begin{figure}
\plotone{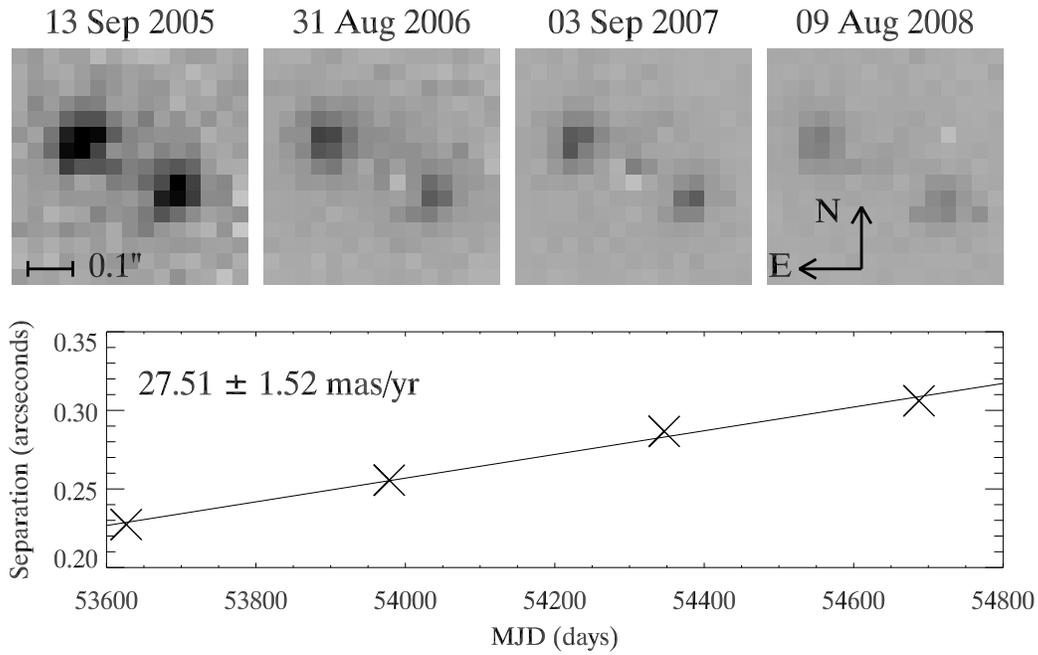}
\caption{Expansion of the torus of emission is shown in both images
and a plot.  The upper panel shows the zero velocity images of
[\ion{Si}{6}] (as in Figure~\ref{fig:const_vel}) at each of our four epochs.
The continuum has been removed from each image.  Images are normalized
and displayed with the same linear stretch.  The 
lower panel plots the measured separation of the two emission peaks as
a function of time.  Each year, the torus expands just less than one
35~mas spaxel.\label{fig:expand}}
\end{figure}

The high spatial resolution of the IFS data allow a velocity
measurement to be made at precise positions of the expanding torus of
material.  We measure the radial velocity along a line perpendicular
to the line defined by the two zero velocity image centroids (see
the solid white line in the upper panel of
Figure~\ref{fig:const_vel}).  The radial velocity line is nearly
coincident to the projection of the polar nodules onto the plane of
the sky so care must be taken when extracting the spectrum.  Here, the
3D data cube allows us to separate the emission in velocity space. As
seen in Figure~\ref{fig:inset}, the northwest polar nodule is more
blueshifted, while the northwest portion of the equatorial torus is
more redshifted.  In a similar way, the southeast polar nodule is more
redshifted, while the southeast portion of the equatorial torus is more
blueshifted.  We extract only the spectrum arising from the equatorial 
torus for our distance determination.     

Our extracted [\ion{Si}{6}] spectrum is double--peaked as shown in
Figure~\ref{fig:const_vel}.  Because we measure the angular expansion 
rate of the
two emission peaks of the zero velocity image from each other, we must
also measure the total velocity separation, $v_{r}$, of the red-- and
blue--shifted emission peaks for each epoch.  We find 
$v_{r}~=~443.3~\pm~12.0~km~s^{-1}$.  The relation of the 
expansion velocity, $v_{exp}$, to the radial velocity is given by:  

\begin{equation} 
\label{vsini} 
v_{exp} = v_{r}sin~i 
\end{equation}  

\noindent where $i$ is the inclination of the system defined such that
edge--on is $i~=~90\degr$.  We measure the semi--major and semi--minor
axes of an ellipsoidal projection of the circular torus onto the image
plane to determine an inclination of $62\fdg0~\pm~1\fdg5$.
Figure~\ref{fig:inset} shows an ellipse graphic using the above
parameters overlaid on the [\ion{Ca}{8}] image taken in September
2007.  By Equation~\ref{vsini}, $v_{exp}~=~502.1~\pm~13.6~km~s^{-1}$.
The measured 
constant velocity over the 5 years of observations is consistent with 
a freely expanding torus unencumbered by shock encounters with 
pre--outburst or interstellar material.   

Using our measured angular expansion rate and expansion velocity, we
can calculate the distance to V723~Cas via expansion parallax:

\begin{equation}
\label{dist}
d [kpc] = 0.211 \frac{v_{exp} [km~s^{-1}]}{\phi [mas~yr^{-1}]}
\end{equation}

By Equation~\ref{dist}, the distance to V723~Cas is
$3.85^{+0.23}_{-0.21}~$kpc.  We consider 3 main sources of errors due to:
1) uncertainty in the angular expansion rate, 2) line fitting, and 
3) inclination uncertainty.  The error in the angular expansion rate 
accounts for most of the error in distance ($^{+0.22}_{-0.20}~kpc$) 
while the other two sources of error account for less than $\pm~0.01~kpc$ 
each.  A summary of data used to determine the expansion parallax is found 
in Table~\ref{tab:expand}.

By comparison, distance estimates vary widely in the literature.  
\citet{evans03} estimate the distance to be 4.0~kpc
employing a combination of methods that yield a range of distances from 
$3.5$ to $4.2~kpc$.  \citet{ness08} assume that the absolute magnitudes of 
HR~Del and V723~Cas are identical to find a distance, then average this 
with other distances in the literature to yield $2.7^{+0.4}_{-0.3}~$kpc.  
The expansion parallax method in this study provides 
the most accurate determination of the distance to V723~Cas.  This distance 
may help provide an improvement in 
the calibration the MMRD relationship for slow novae.  Our distance 
determination of 3.85~kpc is also consistent with the WD mass estimate 
from \citet{evans03} of 0.67~M$_{\sun}$. 
  
\subsubsection{3D Projection} 
\label{sec:3d}
The OSIRIS IFS 3D data cube is a measurement of intensity as a
function of spatial extent and wavelength.  The 3 axes of the cube can
be converted into units of distance in order to visualize the
true shape of the nova ejecta.  The units of the two spatial
dimensions are converted from angle on the sky to length
based on the measured distance to
V723~Cas.  The extent of the nova shell in the line of sight dimension 
can be computed from its expansion velocity.  For our 3D spatial cubes, 
we chose 25 spatial lenslets per side.  At a distance of 3.85~kpc, this 
gives 3370~AU.  For our visualization, we elected not to interpolate over 
spectral pixels such that we rounded the calculated number of spectral 
channels in 3370~AU down to the nearest even integer.  
Table~\ref{tab:cube} shows
this conversion in units of number of spectral channels in 3370~AU for a given
spectral feature and date.  The conversion factor decreases with time
since the measured velocity remains constant as the distance traveled
by the torus increases from year to year.   

\begin{figure}
\plotone{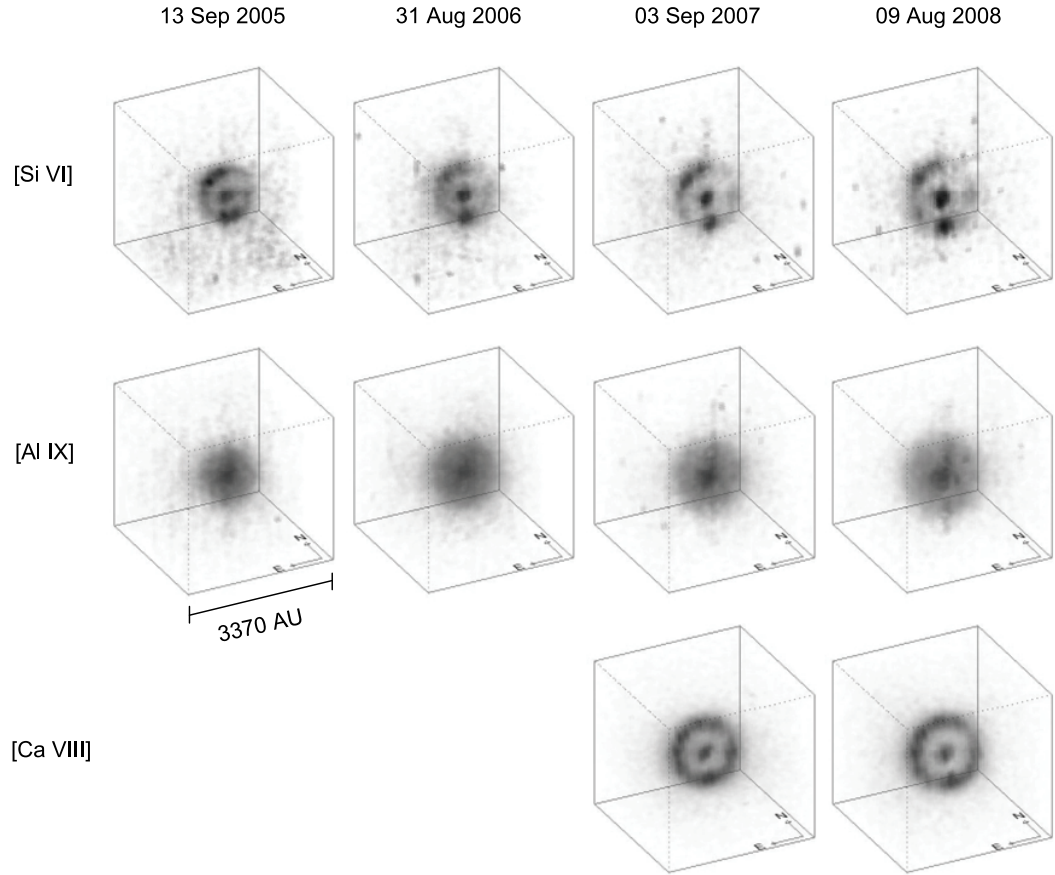}
\caption{Face--on view of V723~Cas in various emission features
after the data were remapped to 3 spatial dimensions.    
The cubes have been rotated as described in \S\ref{sec:3d} and are
3370~AU on a side.  The solid lines
of the cube outline are in the foreground, the dotted lines are in the
background.  Both [\ion{Si}{6}] and [\ion{Ca}{8}] show a distinct torus
of emission whereas [\ion{Al}{9}] has a filled--in shell morphology.  
In all cubes, the bright spot in the center is due to the polar 
emission.\label{fig:face}}
\end{figure}

\begin{figure}
\plotone{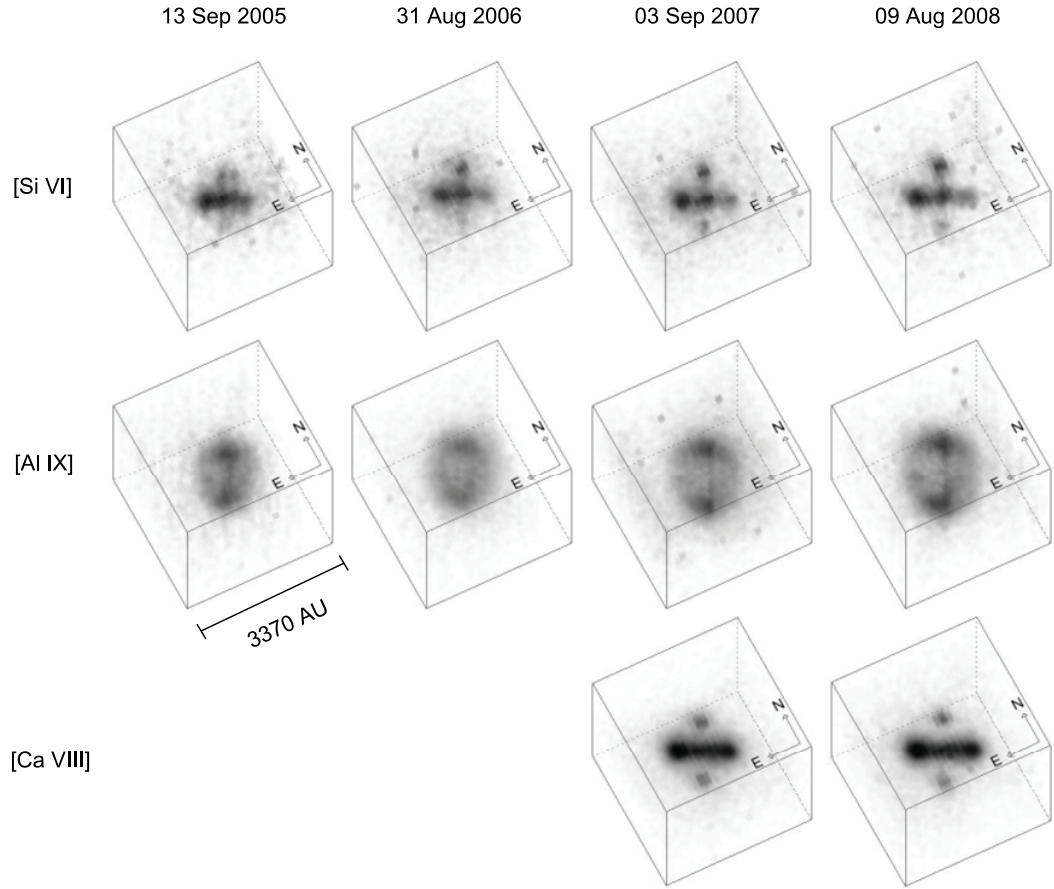}
\caption{Edge--on view of V723~Cas in various emission features
after the data were remapped to 3 spatial dimensions.    
The cubes have been rotated as described in \S\ref{sec:3d} and are
3370~AU on a side.  The solid lines
of the cube outline are in the foreground, the dotted lines are in the
background.  Here, the torus of emission of [\ion{Si}{6}] and
[\ion{Ca}{8}] is clearly separated from the polar emission.  The
[\ion{Al}{9}] polar emission is clearly seen.\label{fig:edge}}
\end{figure}

A data cube of 3 spatial dimensions is rendered for the brightest
spectral features using the volume rendering capabilities of IDL in
Figure~\ref{fig:face} and Figure~\ref{fig:edge}.  The cubes are 3370
AU per side and have a unique linear stretch to accentuate
details; however, all features are fading with time.  As noted earlier,
the continuum from the central source has been removed.
Once we have remapped the data cube to spatial 3D, we can
rotate the cube to show the equatorial and polar emission more clearly.
We start with the image projection of the data cube such that $x$ 
corresponds to east, $y$ to north, and $z$ (formerly $\lambda$) into the 
page.  First we rotate the cube about the $z$ axis by the complement of 
the position angle ($26\fdg3$) such that the projected major axis of the 
equatorial torus is now horizontal.  Next we define a modified $x$ axis 
($x^{\prime}$) along the projected major axis of the equatorial torus.  To 
get the face--on view (Figure~\ref{fig:face}), we rotate about the 
$x^{\prime}$ axis by the inclination 
of $62\degr$.  To get the edge--on view (Figure~\ref{fig:edge}), we 
rotate about the $x^{\prime}$ axis 
in the opposite direction by the complement of the inclination ($-28\degr$).

In the face--on view (Figure~\ref{fig:face}), the torus structure
is clearly visible and we can see knots in the torus in [\ion{Si}{6}] and 
[\ion{Ca}{8}].  The circular
torus of emission confirms the validity of our visualization technique
as our distance determination assumed a circular torus.  The 3D visualization
emphasizes the dramatic difference between the structure of
[\ion{Al}{9}] feature, a prolate spheroid, and the other features,
an equatorial torus with polar nodules.  The bright spot near the
center of each cube is the polar emission.  As the features fade,
residuals of the continuum subtraction become visible as a faint
vertical line of emission passing through the center of the nebula.

In the edge--on view (Figure~\ref{fig:edge}), again we see striking 
differences in the morphologies of
the features.  The polar nodules are visible in all cubes while any
sign of an equatorial torus is absent from the [\ion{Al}{9}] cubes.  

The spatial 3D rendering allows the direct measurement of the 
polar--to--equatorial axial
ratio of the nebula without dependence on viewing angle. Based on data
from 2008, the polar regions were separated by 1670 AU while the
equatorial region spanned only 1270 AU. Thus, the axial ratio is relatively 
small, $1.32~\pm~0.05$.     

\section{Discussion}
\label{discussion}
Morphological differences are not new in nova ejecta.  Using archival
plates of DQ~Her, \citet{mustel70} found an equatorial band and polar
condensations in emission attributed to H$\alpha$ while emission 
attributed to [\ion{O}{3}] showed no equatorial feature.  \citet{harman03} 
resolved an hourglass shaped shell in HR~Del with an equatorial torus
using HST narrow band imaging.  They also found morphological differences
between H and [\ion{O}{3}] emission and point out the
observations can critically constrain our knowledge of the history of
convection during the TNR.   In addition to shell morphology differences
between emission lines, HR~Del has much in common with V723~Cas in 
terms of speed class, orbital parameters, and coronal emission.  One 
difference is the polar--to--equatorial axial ratio of the shell.  
HR~Del has a large axial ratio 
of $1.75~\pm~0.15$ for the prolate
ellipsoidal shell that supports the shell shape rate of decline 
(SSRD) relationship proposed by \citet{slavin95}.  However, the SSRD 
relationship does not hold as strongly for V723~Cas, one of the slowest 
novae on 
record, with its smaller axial ratio of $1.32~\pm~0.05$.  We note that axial 
ratio determined from direct imaging is highly dependent on the 
viewing angle and that future studies with IFS instruments 
could more accurately determine any relation of shell shape with other 
nova parameters. 

There are several possible explanations for the morphological difference
between [\ion{Al}{9}] and [\ion{Si}{6}] and [\ion{Ca}{8}].  The first
is the shell is chemically homogenous, but contains "clumps" 
of denser material embedded within a sparser medium.  This 
interpretation is typically used to resolve how emission from low-- and 
high--ionization potential species can co--exist 
\citep[see $e.g.$][]{lyke01,lyke03}.  
While we do not see uniformly distributed clumps of dense material, 
the equatorial torus could provide some photon shielding.  In fact, the 
ionization potentials of [\ion{Ca}{8}] and [\ion{Si}{6}] are lower than that
of [\ion{Al}{9}] ($147~eV$ and $205~eV$ versus $330~eV$ respectively), but
we do not see an inner equatorial torus of [\ion{Al}{9}] as one might expect
from this interpretation.

A second description may be that the ejecta are inherently homogeneous but 
there are two spatially distinct populations of electrons that permit 
lines from different ions 
to dominate the emission.  Due to the limited wavelength range of
OSIRIS and the resulting lack of the 
necessary combination of emission lines, we are unable to determine the 
electron temperatures or densities within the distinct regions of the 
V723~Cas ejecta.  Previous studies and our NIRSPEC spectra presented here
lacked the spatial resolution to differentiate
between emission regions; thus, any determinations of electron properties
are averaged over the entire shell.  For these reasons, we are unable to 
adequately test the existence of multiple electron populations.

The most intriguing interpretation considers that the shell may not be
homogeneous and the observed differences are a result of
separate ejection events, perhaps from different regions of the
WD.  \citet{lynch08} explain the pronounced second brightening of the 
recent nova V2362~Cyg as a separate TNR ejection event and \citet{evans03} 
suggested that the multiple peaks in the visual light curve of V723~Cas
may be the result of multiple mass ejection events.  If so, multiple mass
ejection events may be common because many novae 
have had multiple peaks in the visual light curves during their early 
evolution.  A possible cause of multiple 
TNR ejection events may be that the TNR does not occur uniformly over the 
surface of the WD.  

As the H--rich material accretes to the WD via a disk near the orbital plane,
it is likely the accreted material is concentrated near the equator of the WD.
Therefore, it is reasonable to assume that the TNR occurs first near the 
equatorial regions of the WD.  If so, then the mass is
ejected mostly in the orbital plane and is subjected to shaping during
the common envelope (CE) phase.  The structure of the V723~Cas ejecta 
is similar to the fast nova V1974~Cyg, with its well--defined equatorial 
torus \citep{paresce95}.  The IR spectrum of V1974~Cyg \citep{wagner96}
exhibits many of the same emission features as in V723~Cas, but at
much higher expansion velocities.  \citet{paresce95} discuss the ejecta may
have been shaped by the CE phase with binary orbit
depositing angular momentum into the CE, thereby enhancing the density 
into the
equatorial plane.  If the fast moving ejecta of V1974~Cyg was shaped into an
equatorial torus by its CE phase, it is likely the binary orbit 
shaped the much slower ejecta of V723~Cas during its CE phase.  
The energy from the TNR in the 
equatorial region may heat the higher latitudes of the WD, 
eventually triggering additional TNR events.  If so, the subsequent mass
ejections may be concentrated towards the poles.  Any observations of 
abundance gradients in nova ejecta may help constrain the knowledge of the 
history of convection during the TNR \citep{starrfield08}.  \citet{lloyd97} 
modeled the CE phase and showed 
that slower novae are more likely to have the envelope ejected in the 
plane of the binary orbit.  However they also predict very little mass 
loss in the polar direction.  \citet{porter98} enhanced the CE model of 
\citet{lloyd97} by
including envelope rotation which can produce more prolate shaped ejecta. 
Our observations of V723~Cas should be able to help constrain 
models such as these to learn more about the pre--outburst accretion phase 
and material mixing on the WD. 
 
If an equatorial torus is a result of the CE
phase, the inclination of the torus can be assumed to be the same as the
inclination of the binary orbit.  For V723~Cas, we measure the inclination 
to be $62\fdg0~\pm~1\fdg5$.  Roche geometry cannot produce an eclipse 
light curve at
this inclination \citep{horne85}; thus, the sawtooth pattern in the light 
curve with orbital period of $0\fd693265$ measured by \citet{goranskij07} 
is likely due to heating effects of the secondary surface that faces the 
WD.  We would expect the amplitude of the light curve variations 
in V723~Cas to diminish with time as the radiation from the primary 
returns to quiescence.

\section{Conclusions}  
We have presented IFS observations of V723~Cas with unprecedented spatial
resolution in the IR.  These data allow us to remove the point--like 
continuum and
create custom narrowband "filter" images of coronal emission features 
that allow us to see 2 distinct 
ejecta morphologies.  Furthermore, our dataset allows us to determine a 
highly accurate distance of $3.85^{+0.23}_{-0.21}~$kpc.  We have used
this distance to convert our $x-y-\lambda$ data cubes into $x-y-z$ spatial
data cubes that allow us to compare the true shapes various emission lines.  
The morphological differences are likely due to separate ejection events 
possibly due to an uneven TNR combined with some self--shielding of X--rays
by the equatorial torus.  We anticipate that these V723~Cas data and future
IFS data of classical novae will help constrain models that attempt to 
describe the nature of the outburst.

\acknowledgements
The authors wish to thank T.~E. Armandroff and F.~H. Chaffee, current and
past directors of the W.~M.~Keck Observatory respectively, for access 
to their director's time via Team Keck.

The authors wish to thank J.~E. Larkin and S.~A. Wright for valuable
insights into OSIRIS and its DRP and R.~W. Goodrich for constructive
feedback on an early draft of this paper.

The data presented herein were obtained at the W.~M.~Keck 
Observatory, which is operated as a scientific partnership among the 
California Institute of Technology, the University of California and the 
National Aeronautics and Space Administration. The Observatory was made 
possible by the generous financial support of the W.~M.~Keck Foundation.  

The authors wish to recognize and acknowledge the very significant cultural 
role and reverence that the summit of Mauna Kea has always had within the 
indigenous Hawaiian community.  We are most fortunate to have the 
opportunity to conduct observations from this mountain.  

{\it Facilities:} \facility{Keck:II (OSIRIS)}, \facility{Keck:II (LGSAO)},
\facility{Keck:II (NIRSPEC)}

\clearpage 

\newpage  

\begin{deluxetable}{ccccccc}  

\tablecaption{Observational Details\label{tab:details}  } 

\tablehead{ 
\colhead{UT Date}			& 
\colhead{Days since}			& 
\colhead{Instrument} 			& 
\colhead{Filter}			& 
\colhead{$\lambda$ Range}		& 
\colhead{exp time}			& 
\colhead{WFS Rate}			\\ 
\colhead{}				& 
\colhead{outburst\tablenotemark{a}}	& 
\colhead{}				& 
\colhead{}				& 
\colhead{($\micron$)}			& 
\colhead{(s)}				& 
\colhead{(Hz)\tablenotemark{b}}		
}  

\startdata  

2004 Aug 22 & 3285.6 & NIRSPEC & N--3 & 1.1--1.3 & 300 & n/a \\ 
2004 Aug 22 & 3285.6 & NIRSPEC & N--7 & 1.9--2.2 & 300 & n/a \\ 
2004 Aug 22 & 3285.6 & NIRSPEC & N--7 & 2.2--2.6 & 300 & n/a \\ 
2005 Sep 13 & 3672.4 & OSIRIS  & Kn1 & 1.955 -- 2.055 & 900 & 300 \\ 
2005 Sep 13 & 3672.4 & OSIRIS  & Kn2 & 2.036 -- 2.141 & 900 & 300 \\ 
2005 Sep 13 & 3672.4 & OSIRIS  & Kn3 & 2.121 -- 2.229 & 900 & 300 \\ 
2006 Aug 31 & 4024.4 & OSIRIS  & Kn1 & 1.955 -- 2.055 & 1800 & 300 \\ 
2006 Aug 31 & 4024.4 & OSIRIS  & Kn3 & 2.121 -- 2.229 & 1800 & 300 \\ 
2007 Sep 03 & 4392.5 & OSIRIS  & Kn1 & 1.955 -- 2.055 & 3600 & 800 \\ 
2007 Sep 03 & 4392.5 & OSIRIS  & Kn3 & 2.121 -- 2.229 & 3600 & 800 \\ 
2007 Sep 03 & 4392.5 & OSIRIS  & Kn5 & 2.292 -- 2.408 & 1800 & 800 \\ 
2008 Aug 09 & 4733.5 & OSIRIS  & Kn1 & 1.955 -- 2.055 & 5400 & 800 \\ 
2008 Aug 09 & 4733.5 & OSIRIS  & Kn3 & 2.121 -- 2.229 & 3600 & 800 \\ 
2008 Aug 09 & 4733.5 & OSIRIS  & Kn5 & 2.292 -- 2.408 & 1800 & 800 \\  

 \enddata 
\tablenotetext{a}{Discovery date is August 24.5 1995 (JD 2,449,954)} 
\tablenotetext{b}{WFS is wavefront sensor} 
\end{deluxetable}  

\clearpage 

\newpage  

\begin{deluxetable}{ccccccc}  

\tablecaption{Relative Fluxes\label{tab:linelist}  } 

\tablehead{ 
\colhead{Wavelength}			& 
\colhead{}				& 
\multicolumn{5}{c}{F/F([\ion{Si}{6}])\tablenotemark{a}} \\ 
\cline{3-7}				\\
\colhead{($\micron$)}			& 
\colhead{Identification}		& 
\colhead{2004}				& 
\colhead{2005}				& 
\colhead{2006}				& 
\colhead{2007}				& 
\colhead{2008}		
}  

\startdata  

1.87561 & Pa$\alpha$ & 1.596 & \nodata &  \nodata &  \nodata & \nodata \\
1.94509 & Br8 & 0.194 & \nodata & \nodata &  \nodata & \nodata \\
1.96287 & [\ion{Si}{6}] & 1.000 & 1.000 & 1.000 & 1.000 & 1.000 \\
2.03788 & \ion{He}{2} 15--8 & \nodata & 0.065 & \nodata &  \nodata & \nodata \\
2.0443  & [\ion{Al}{9}] & 0.342 & 0.786 & 1.022 & 1.106 & 1.213 \\
2.0581  & \ion{He}{1} & 0.017 & \nodata & \nodata & \nodata & \nodata \\
2.0996  & ? & 0.022 & 0.062 & \nodata & \nodata & \nodata \\
2.1156  & \ion{He}{1} & 0.015 & \nodata & \nodata & \nodata & \nodata \\
2.16553 & Br$\gamma$ & 0.091 & 0.390 & 0.381 & 0.560 & 0.886 \\
2.18911 & \ion{He}{2} 10--7 & 0.048 & 0.142 & 0.221 & 0.248 & 0.388 \\
2.206   & [\ion{Ti}{7}] & 0.027 & 0.067 & 0.049 & 0.099 & 0.071 \\
2.2190  & ? & 0.050 & 0.097 & 0.165 & 0.185 & 0.142 \\
2.32141 & [\ion{Ca}{8}] & 3.200 & \nodata & \nodata & 5.519 & 6.144 \\
2.34704 & \ion{He}{2} 13--8 & 0.021 & \nodata & \nodata & \nodata & \nodata \\
2.4807  & [\ion{Si}{7}] & 4.111 & \nodata & \nodata & \nodata & \nodata \\

\enddata 
\tablenotetext{a}{V723~Cas was dimmer each year it was observed.  
The relative flux of [\ion{Si}{6}] is as follows: 2005, 1.000; 2006, 
0.614; 2007, 0.367; 2008, 0.271.} 
\end{deluxetable}  

\clearpage 

\newpage  

\begin{deluxetable}{ccccc}  

\tablecaption{[\ion{Si}{6}] Expansion Parameters\label{tab:expand}  } 

\tablehead{ 
\colhead{UT Date}			& 
\colhead{MJD}				& 
\colhead{Separation}			& 
\colhead{Separation}			& 
\colhead{$v_{exp}$}			\\ 
\colhead{}				& 
\colhead{}				& 
\colhead{($\arcsec$)}			& 
\colhead{(AU)\tablenotemark{a}}		& 
\colhead{($km~s^{-1}$)\tablenotemark{b}} 
}
  
\startdata  

2005 Sep 13 & 53626.4 & 0.2273 & 875 & 489.5 \\ 
2006 Aug 31 & 53978.4 & 0.2556 & 984 & 491.3 \\ 
2007 Sep 03 & 54346.5 & 0.2866 & 1103 & 512.0 \\ 
2008 Aug 09 & 54687.5 & 0.3061 & 1178 & 515.5 \\  

\enddata 
\tablenotetext{a}{Using d~=~3.85~kpc (see \S\ref{distance})} 
\tablenotetext{b}{Corrected for an inclination of $62\degr$} 
\end{deluxetable}  

\clearpage 

\newpage  

\begin{deluxetable}{cccccc}  

\tablecaption{Data Cube Spectral Channel to Spatial Extent 
Conversion\tablenotemark{a}\label{tab:cube}  } 

\tablehead{ 
\colhead{}				& 
\colhead{Feature}			& 
\colhead{2005}				& 
\colhead{2006}				& 
\colhead{2007}				& 
\colhead{2008}				
}

\startdata  

& [\ion{Si}{6}] & 46 & 42 & 38 & 34 \\ 
& [\ion{Al}{9}] & 48 & 44 & 40 & 36 \\ 
& [\ion{Ca}{8}] & \nodata & \nodata & 46 & 42 \\ 

\enddata 
\tablenotetext{a}{Number of spectral channels that approximate 3370~AU} 

\end{deluxetable}  

\end{document}